\documentclass[aps,prd,twocolumn,superscriptaddress,amsmath,showpacs]{revtex4}

\newcommand{\e}{\epsilon}
\newcommand{\Ft}{\widetilde{F}}
\newcommand{\m}{\mu} 
\newcommand{\n}{\nu} 
\newcommand{\la}{\lambda}
\newcommand{\ka}{\kappa}
\newcommand{\p}{\partial}  
\newcommand{\be}{\begin{equation}}
\newcommand{\ee}{\end{equation}}
\newcommand{\de}{\delta}
\newcommand{\tht}{\theta}
\newcommand{\si}{\sigma}
\newcommand{\g}{\gamma}
\newcommand{\na}{\nabla}

\begin{document}
\preprint{CBPF-NF-018/04}
\preprint{hep-th/0407109}

\title{Planar Supersymmetric Quantum Mechanics of a Charged Particle in an External Electromagnetic Field}
%\footnote{A previous version of this work was presented in poster format by JAHN during the II International 
%Conference on Fundamental Interactions, June 2004, Pedra Azul-ES, Brazil.} 
%
\author{Ricardo C. Paschoal}\email{paschoal@cbpf.br}
\affiliation{Centro Brasileiro de Pesquisas F\'{\i}sicas -- CBPF, \\
Rua Dr. Xavier Sigaud 150, 22290-180, Rio de Janeiro, RJ, Brasil}
\affiliation{Servi\c{c}o Nacional de Aprendizagem Industrial,  \\
Centro de Tecnologia da Ind\'{u}stria Qu\'{\i}mica e T\^{e}xtil -- SENAI/CETIQT,  \\
Rua Dr. Manoel Cotrim 195, 20961-040, Rio de Janeiro, RJ, Brasil}
\author{Jos\'{e} A. Helay\"{e}l-Neto}\email{helayel@cbpf.br}
\affiliation{Centro Brasileiro de Pesquisas F\'{\i}sicas -- CBPF, \\
Rua Dr. Xavier Sigaud 150, 22290-180, Rio de Janeiro, RJ, Brasil}
\affiliation{Grupo de F\'{\i}sica Te\'{o}rica Jos\'e Leite Lopes, \\
P.O.\ Box 91933, 25685-970, Petr\'opolis, RJ, Brasil}
\author{Leonardo P. G. de Assis}\email{lpgassis@cbpf.br}
\affiliation{Centro Brasileiro de Pesquisas F\'{\i}sicas -- CBPF, \\
Rua Dr. Xavier Sigaud 150, 22290-180, Rio de Janeiro, RJ, Brasil}
\affiliation{Grupo de F\'{\i}sica Te\'{o}rica Jos\'e Leite Lopes, \\
P.O.\ Box 91933, 25685-970, Petr\'opolis, RJ, Brasil}
% Please use the appropriate macro for each type of information

%\author{}
%\thanks{}
%\affiliation{}


\date{\today}

\begin{abstract}
The supersymmetric quantum mechanics of a two-dimensional non-relativistic particle subject to external magnetic and  electric fields is studied in a superfield formulation and with the typical non-minimal coupling of (2+1) dimensions.
Both the N=1 and N=2 cases are contemplated and the introduction of the electric interaction is suitably analysed.
\end{abstract}

%\keywords{}

\maketitle

% If in two-column mode, this environment will change to single-column
% format so that long equations can be displayed. Use
%\begin{widetext}
%\end{widetext}

\section*{\label{int}Introduction}

Since the pioneering papers on supersymmetric quantum mechanics (SQM)\cite{Nicolai,Witten81,Witten82}, a great deal of work on the subject has been done, including various reviews\cite{SQMreview0,SQMreview1,SQMreview2,SQMreview3,SQMreview4,SQMreview5} and
books\cite{SQMbook0,SQMbook1,SQMbook2,SQMbook3,SQMbook4}, the research in the field being still active.
In particular, a very usual 
question in this field
is the realization 
of supersymmetry (SUSY)
in quantum-mechanical systems
involving 
charged or neutral particles in interaction with magnetic fields, in various space dimensionalities. Not related to SQM, however, it is a well-known fact that in (2+1) dimensions a non-minimal coupling naturally arises\cite{Stern91,Kogan91,Lat-Soro-91,Lat-Soro-91b} and 
allows for a magnetic moment interaction even in the case of spin-zero 
particles (scalar matter fields). These two aspects, SQM and non-minimal coupling, have not yet
been contemplated simultaneously in the literature, and so the present work is intended to address this problem.

Here,
from the very beginning, a superfield formulation is carried out that involves charged particles 
with magnetic moment subject to external electric and magnetic fields whose potentials are 
functions of the particle superfield coordinates. Both N=1 
and N=2 cases are considered.

Another interesting question that remains open in the literature is whether an electric field interaction may be present without explicitly breaking SUSY. For N=1-SQM, the traditional answer is no\cite[p.51]{SQMbook1}, but here this 
question is also reassessed and it is shown that, in a non-minimal coupling scheme, this indeed may 
occur: an N=1 supersymmetric quantum-mechanical system is proposed, where the electric 
field interaction appears along with the coupling between the magnetic dipole moment and the magnetic
field. It is shown under which conditions this may take place. In the case N=2, Witten's model\cite{Witten81,Witten82} is the most celebrated and the one with more applications. The corresponding literature shows that an electric interaction (via a scalar potential) is possible within such supersymmetric models, but it occurs only in each of the two sectors (`bosonic' and `fermionic') of the Hamiltonian: the two electric potentials (the `bosonic' and the `fermionic'), although deriving from the same superpotential, have different expressions in terms of it and thus do not refer simultaneously to the same particle, but rather refer to two almost isospectral systems (the `almost' here refers to the ground state), typical of (unbroken) supersymmetric systems. On the other hand, in the N=2-(N=1-)SUSY of Pauli equation in two (three) space dimensions\cite{SQMbook2,g2sqm}, the two sectors of the Hamiltonian (the `bosonic' and the `fermionic' ones) refer to the two different spin states of the \emph{same}\/ spin-$1/2$ system. In the present work, a proposition is made about the possibility of a supersymmetric Pauli Hamiltonian in (2+1) dimensions including electric interactions, with a non-minimal coupling.

The outline of the present paper is as follows. In Section~\ref{nmin}, a brief review of the (2+1)-dimensional non-minimal coupling is presented. Next, N=1- and N=2-SQM are discussed in Sections~\ref{N=1} and \ref{N=2}, respectively. Finally, in Section~\ref{concl}, the General Conclusions are drawn.

\section{\label{nmin}Non-minimal coupling in (2+1) dimensions}

In (3+1) dimensions the dual $\Ft^{\m\n} \equiv \frac{1}{2}\e^{\m\n\ka\la}F_{\ka\la}$ of the electromagnetic field,  $F_{\ka\la}$, is a second-rank tensor. On the other hand, in (2+1) dimensions, it is a vector, $\Ft^\m \equiv \frac{1}{2}\e^{\m\ka\la}F_{\ka\la}$, and, as shown in Refs.~\cite{Stern91,Kogan91,Lat-Soro-91,Lat-Soro-91b}, it is possible to define a non-minimal derivative,   
\be
        {\cal D}_\m  \equiv  \p_\m + iqA_\m + ig\Ft_{\m}  , \label{cov-deriv}
\ee
where $g$ is the planar analogue of the magnetic dipole moment, which couples non-minimally to the magnetic field. 

Such a derivative implies that the term 
\be
q\Phi -gB \label{add-U}
\ee
must be added to the Schr\"odinger equation for an electron subject to an electromagnetic field. Also, the momentum, $\vec p=-i\vec\nabla$, must be replaced with
\be
\vec p-q\vec A+g\widetilde{\vec{E}} \label{add-p}.
\ee

These substitutions are readily seen as equivalent to the minimal prescription, except for the following changes:
\be
       \Phi \rightarrow \Phi' = \Phi -\frac{g}{q}B \label{redef-Phi}
\ee
and
\be
       (\vec{A})_i \rightarrow (\vec{A}')_i = (\vec{A})_i - \frac{g}{q}\widetilde{E}_i , \label{redef-A}
\ee
or: 
\be
A_\m \rightarrow A'_\m \equiv  A_\m + \frac{g}{q}\Ft_{\m}. \label{red-Amu}
\ee

Equation (\ref{redef-A}) implies that the magnetic field is redefined as:
\be
       B \rightarrow B' = B + \frac{g}{q}\left( \vec{\nabla}\cdot\vec{E} \right) . \label{redefB}
\ee

It is worthwhile to stress here that the field redefinitions above,  
though formally acceptable, should not be used to eliminate physical effects inherent to the non-minimal coupling. Indeed, the latter describes a magnetic dipole moment even for scalar particles and may lead to contact interactions between the test particle and the eventual sources of the electromagnetic field acting upon the particle. The electromagnetic field, though non quantized, is not a fixed background. It is generated by external sources and they may induce interactions once the test particle is located in the region of the sources.

The non-minimal coupling studied here may be considered as resulting from the dimensional reduction of a Lorentz-breaking Chern-Simons model in (3+1) dimensions\cite{CaFiJa90,Belich-2,Belich-3,Belich-5}, defined by the following derivative:
\be
        \p_\m + iqA_\m + i\frac{\g}{2}\e_{\m\n\ka\la}v^\n F^{\ka\la}, \label{Lor-brk}
\ee
where $\g$ is a constant (like $q$, a property of the particle), $\e_{\m\n\ka\la}$ is the (Levi-Civita) totally antisymmetric tensor in (3+1) dimensions and $v^\n$ is a fixed (Lorentz-breaking) vector in spacetime. Indeed, performing the corresponding steps in order to obtain the Schr\"odinger equation for a charged particle, one obtains that it is equivalent to add the term
\be
       q\Phi - \g\vec{v}\cdot\vec{B}
\ee
to the Hamiltonian and substitute the momentum with
\be
      \vec p - q\vec A + \g v^0 \vec B - \g\vec v \times \vec E.
\ee

Thus, choosing $v^\n=(0, \vec v)$ and $\g\vec v = (0,0,\g v^3)$, one immediately verifies that the redefinitions stated in Eqs.~(\ref{add-U}--\ref{add-p}) are exactly recovered, with the (3+1)-dimensional quantity $\g v^3$ playing the role of its (2+1)-dimensional counterpart $g$, and with only the third ($z$) component of the (3+1)-dimensional magnetic field $\vec B$ and the in-plane ($x,y$) components of the (3+1)-dimensional electric field $\vec E$ contributing to the Hamiltonian, just as it should be in (2+1) dimensions.

In connection with planar physics, it is natural to invoke the Quantum Hall Effect (QHE)\cite{FQHE1,FQHE2,FQHE3,FQHE4} and its fractional version (FQHE). Indeed, in Ref.~\cite{PLA}, a parallel was made between the particle with charge $q$ and magnetic moment $g$, as described here, and the composite fermion (CF) of Jain's model for the FQHE\cite{Jain1,Jain2,Jain3,Jain4}, in which $g$ is associated to the CF magnetic flux. However, in terms of a 4-dimensional theory with Lorentz violation\cite{CaFiJa90}, the CF flux, $g$, contains a contribution from the particle itself (by means of the parameter $\g$) and another one from the background vector, $v^\nu$, that breaks Lorentz symmetry. An interpretation is also possible for $\vec v$: it is responsible for the confinement of the electrons to the plane and, therefore, it is natural to relate $\vec v$ to the $z$-component of the three-dimensional magnetic field, which is very large in the FQHE ($\sim$10 T) and forbids the electrons to move in the $z$-direction, breaking in this way their (3+1)-dimensional Lorentz symmetry. Indeed, the construction of a CF is possible only in two space dimensions.  

Now, if one wishes to supersymmetrize this model, it is important to notice that this is not possible 
(in N=1-SQM) for a scalar potential interaction such as the one given by expression (\ref{add-U}). 
Therefore, in order to keep invariance under N=1-SUSY, it is necessary that:
\be
        gB(x, y) = q\Phi(x, y).  \label{Phi-SUSY}
\ee

\section{\label{N=1}N=1-SQM }

A charged planar particle non-minimally coupled to a magnetic field is described as an $N=1$-SQM system by means of the superspace action below:
\be
        S_1 = \frac{iM}{2}\int{dt d\tht (D\vec X)\cdot\dot{\vec{X}}}  +  iq \int{dt d\tht D\vec X\cdot\vec{A}'(\vec X)}, \label{S1}
\ee
where $\vec{A}'(\vec X)$ is the vector (super)potential in a non-minimal coupling scheme, given by Eq.~(\ref{redef-A}), and $\vec X(t)$ is the real ``superfield" (in fact, the supercoordinate of the particle), given by
\be
      X^j(t,\tht) = x^j(t) + i\tht\la^j(t),\;\;\;\; j = (1,2),     \label{X}
\ee
$x^j(t)$ being the two real coordinates of the planar particle, $\la^j(t)$ their Grassmannian supersymmetric partners and $\tht$ the real, Grassmannian supersymmetric coordinate that parametrizes the superspace, $(t;\tht)$.
The supersymmetry covariant derivative $D$ is given by:
\be
      D = \p_\tht - i\tht\p_t .
\ee

The action (\ref{S1}), $S_1 = \int{dt L_1}\:$, reads in components of the superfield as:
\begin{eqnarray}
      L_1 & = & \frac{M{\dot{\vec{x}}}^2}{2} - \frac{iM}{2}\dot{\vec\la}\cdot\vec\la + q\dot{\vec x}\cdot\vec A 
                      - g\dot{\vec x}\cdot\widetilde{\vec E} + \nonumber \\*
              &   &  {} - \frac{iq}{2}(\vec{\la}\times\vec{\la})B - \frac{ig}{2}(\vec{\la}\times\vec{\la})(\vec\nabla\cdot\vec E),  
\end{eqnarray}
where one notices that
\be
        \vec{\la}\times\vec{\la} \equiv \e_{ij}\la_i\la_j = [\la_1, \la_2].
\ee

A convenient change of variables will be performed:
\begin{eqnarray}
      \psi \equiv \sqrt{\frac{M}{2}}(\la_1 + i \la_2) \\
      \bar\psi \equiv \sqrt{\frac{M}{2}}(\la_1 - i \la_2),
\end{eqnarray}
giving rise to the following expression for the Lagrangian:
\begin{eqnarray}
     L_1 & = &  \frac{M{\dot{\vec{x}}}^2}{2} - \frac{i}{2}(\dot{\psi}\bar\psi + \dot{\bar\psi}\psi) + q\dot{\vec x}\cdot\vec A 
                     - g\dot{\vec x}\cdot\widetilde{\vec E} + \nonumber \\*
            &    & {}+ \frac{q}{2M}[\psi,\bar\psi]B + \frac{g}{2M}[\psi,\bar\psi](\vec\nabla\cdot\vec E).  \label{L1}
\end{eqnarray}

The corresponding Hamiltonian will be obtained after a canonical quantization procedure following Ref.~\cite[p.46]{SQMbook2}.
The Grassmannian momenta are defined as
\begin{eqnarray}
      \pi \equiv \frac{\p L_1}{\p\dot{\psi}} = -\frac{i}{2}\bar\psi             \\
      \bar\pi \equiv \frac{\p L_1}{\p\dot{\bar\psi}} = -\frac{i}{2}\psi,
\end{eqnarray}
leading 
to the following operator algebra ($\hbar=1$):
\begin{eqnarray}
                        [ x_i, p_j ] = i\de_{ij}, & \;\;\;\;\;\; & \{ \psi, \pi \} = \{ \bar\psi, \bar\pi \} = - \frac{i}{2} \\
                        \{ \psi, \bar\psi \} = 1, & \;\;\;\;\;\; & \{ \pi, \bar\pi \} = - \frac{1}{4},
\end{eqnarray}
besides $\pi^2=\bar\pi^2=\psi^2=\bar\psi^2=0$. These relations may be represented by
\begin{eqnarray}
                        \psi = \si_+, & \;\;\;\;\;\; & \bar\psi = \si_-  \label{rep1} \\
                        \pi = -\frac{i}{2}\si_-, & \;\;\;\;\;\; & \bar\pi = -\frac{i}{2}\si_+, \label{rep2}
\end{eqnarray}
where the $\si$'s are the Pauli matrices (and there is no other inequivalent representation\cite{LaMi89}).

The quantized version of the Hamiltonian is:
\be
H_1 = \frac{\left(\vec p - q\vec A + g\widetilde{\vec E}\right)^2}{2M} - \frac{qB}{2M}\si_3 -
              \frac{g(\vec\nabla\cdot\vec E)}{2M}\si_3,          \label{H1Tq}
\ee
where the relation $[\si_+, \si_-]=\si_3$ was used. This Hamiltonian automatically reveals a spin-1/2 particle with magnetic dipole moment $q\si_3/2M$ and gyromagnetic ratio 2, as expected, and in agreement with Ref.~\cite{g2sqm}, about SQM (but without the superfield formulation used here), and 
Refs.~\cite{g2anyon-1,g2anyon-2,g2anyon-3,g2anyon-4,g2anyon-5}, with general arguments concerning particles in (2+1) dimensions.

It is interesting to compare this Hamiltonian with the one obtained in Ref.~\cite{PLA}, as the non-relativistic limit of the non-minimal (2+1)-dimensional Dirac equation:
\be
      H =
       q\Phi + \frac{ \left( \vec p - q\vec{A} + g\widetilde{\vec{E}} \right)^2 }{2M} - \frac{qB}{2M} - gB 
      - \frac{g}{2M}\left( \vec{\nabla}\cdot\vec{E} \right)      .      
      \label{2+1Dir-nrel-nmin}       
\ee

As already mentioned above, the condition $gB = q\Phi $ is necessary in order to keep the N=1-SUSY. Thus, under such a condition, 
\be
      H =
        \frac{ \left( \vec p - q\vec{A} + g\widetilde{\vec{E}} \right)^2 }{2M} - \frac{qB}{2M} 
      - \frac{g}{2M}\left( \vec{\nabla}\cdot\vec{E} \right)      .      
      \label{2+1Dir-nrel-nmin-no-phi}       
\ee

Comparing Eqs.~(\ref{H1Tq}) and (\ref{2+1Dir-nrel-nmin-no-phi}), one concludes that the spin-up component of the former 
equals the latter. The same occurs with the spin-down component  when a representation different from Eqs.~(\ref{rep1}--\ref{rep2}) is used, in which the matrices $\si_+$ and $\si_-$ are interchanged.

The last term in Eq.~(\ref{H1Tq}) may be related to the magnetic field, 
in the case of Maxwell-Chern-Simons (MCS) theory\cite{PLA,DJT82}, in which the following field equations hold:
\begin{eqnarray}
 & & \vec{\nabla}\cdot\vec{E}-m_{cs}B=\rho \label{MCS1} \\ 
 & & \vec{\nabla}\times\vec{E}= - \frac{\p B}{\p t} \label{MCS2} \\
 & & \widetilde{\vec{\nabla}}B - m_{cs}\widetilde{\vec{E}} = \vec{J} +  \frac{\p \vec{E}}{\p t}\label{MCS3},
\end{eqnarray}
where, as above, $ \widetilde{\vec{\nabla}}_i  \equiv  \e_{ij}\p_j $, 
and $m_{cs}$ is the Chern-Simons (topological) mass parameter, the (gauge-symmetry preserving) mass of the gauge field. Indeed, in the region outside external charges ($\rho=0$), the Hamiltonian (\ref{H1Tq}) turns into the following expression:
\be
     H_1 = \frac{\left(\vec p - q\vec A + g\widetilde{\vec E}\right)^2}{2M} - \frac{q\si_3}{2M}\left( 1 + \frac{gm_{cs}}{q} \right)B.
               \label{H1Tq-MCS1}
\ee

From this Hamiltonian, it is natural to define an \emph{effective gyromagnetic ratio}, 
%$\gamma_{\mbox{\scriptsize eff}}$
$\gamma_{\text{eff}}$, whose departure from 2 is given by
\be
%       \gamma_{\mbox{\scriptsize eff}} - 2 = \frac{gm_{cs}}{q} - 1,
          \gamma_{\text{eff}} - 2 = \frac{gm_{cs}}{q} - 1,
\ee  
which reinforces the well-known fact that $g$ is to be interpreted as an \emph{anomalous}\/ magnetic dipole moment. In this context, the condition
\be
       gm_{cs}/q=1 \label{g-crit}
\ee
is necessary in order to keep the effective gyromagnetic ratio in its standard value 2. Interestingly, such a condition was also obtained in field theoretical works, with other interpretations: it turns interacting MCS theory into a free one and relates it to pure-CS theory and anyons\cite{Stern91,GeWa92}; it gives rise to no one-loop radiative corrections to the photon mass\cite{GeWa92}; and it reduces the differential equations for the gauge fields from second- to first-order, allowing one to get vortex solutions\cite{Tor92}.

\section{\label{N=2}N=2-SQM}
The superfield formulation of Witten's (one space dimension, N=2-) SQM may be found in Refs.~\cite{SQMbook4,Coop-Freed,Lanc}, in terms of a scalar superpotential (a function of the one-dimensional real supercoordinate). A generalization to $d$ space dimensions is presented in Refs.~\cite{SQMbook1,Lanc}, also in terms of a scalar superpotential, but now as a function of $d$ real superfield coordinates. A different approach to two space dimensions, using a vector superpotential instead of a scalar one, is outlined in Ref.~\cite{Das}, but without a superfield formulation. 

The N=2-SQM of Pauli equation in two space dimensions is formulated in terms of (complex) chiral and anti-chiral superfield coordinates in Refs.~\cite{Lanc,CLN-1,CLN-2}, by means of a (K\"ahler) super(pre)potential (a function of those superfield coordinates).
The introduction of an electric interaction into the planar Pauli equation without the explicit breaking of SUSY was made in Ref.~\cite{Tkachuk}, but there a non-stationary magnetic field was considered. In Refs.~\cite{Iof-1,Iof-2,Iof-3}, the Pauli operator (including an external scalar potential) in two space dimensions is identified with the 2$\times$2 component of a total 4$\times$4 super-Hamiltonian. An N=2-superfield formulation encompassing all these issues, viz., Pauli equation in (2+1) dimensions with electric interactions, and also considering the planar non-minimal coupling studied in Section~\ref{nmin}, is lacking. The present Section is devoted to fill this gap.
Non-stationary situations are not considered in this paper, and so the electric interaction is due only to a scalar potential. Also, the mentioned possibility of the Pauli operator to be a component of the total super-Hamiltonian will not be considered here, but rather it will always be regarded as the total super-Hamiltonian itself.

It was shown in Section~\ref{nmin} that, in order to obtain the Schr\"odinger equation with a non-minimal coupling, it is necessary to add the term (\ref{add-U}) to the free Hamiltonian, and also to perform the replacement expressed by Eq.~(\ref{redef-A}). If condition (\ref{Phi-SUSY}) is valid, then there is no scalar potential interaction in the resulting Hamiltonian, which therefore becomes `pure-magnetic', allowing one to derive it from the chiral superaction of Ref.~\cite{CLN-1}.  Such a superaction contains, instead of $\vec A(x,y)$, the (real) K\"ahler prepotential $K(x,y)$ (as will be seen below), which satisfies the following relations (from now on, $A_i$ stands for $(\vec A)_i$):
\begin{eqnarray}
                             A_j & = & \e_{jk}\p_kK \\
                             B\equiv\vec\na\times\vec A & \equiv & \e_{ij}\p_i A_j=-\na^2K. \label{B-K}
\end{eqnarray}
Therefore, it would be desirable to find out how to implement the non-minimal prescription of Eq.~(\ref{redef-A}) in terms of the prepotential $K(x,y)$. This is done as follows:
\begin{eqnarray}
                              A'_i & = &  A_i - \frac{g}{q}\widetilde{E}_i  \nonumber \\
		             & = &  \e_{ij}\p_jK - \frac{g}{q}\e_{ij}E_j \equiv \e_{ij}\p_jK - \alpha\e_{ij}E_j  \nonumber \\
                                     & = &  \e_{ij}\left[ \p_jK + \alpha\left(\p_tA_j + \p_j\Phi\right) \right] \nonumber \\
			 & = &  \e_{ij}\p_j\left( K + \alpha\Phi  \right) \equiv \e_{ij}\p_jK',
\end{eqnarray}
where the stationary condition $\p_t=0$ was used. Thus, the required prescription may be considered as:
\be
       K\rightarrow K' = K + \alpha\Phi = K + \alpha^2B = K - \alpha^2\na^2K. \label{redef-K}
\ee

Turning now to the chiral superaction, and using a notation  similar to that of Ref.~\cite{CLN-1}, the superspace coordinates are the time, $t$, and the Grassmanian variables, $\tht$ and $\bar\tht$ (the bar over a quantity stands for its complex or Hermitian conjugate). In a non-minimal coupling scheme, the N=2 superaction for a planar particle with mass $M$ and electric charge $q$ in a magnetic field satisfying Eq.~(\ref{B-K}) is given by:
\be
       S_2 = \frac{M}{8}\int{dt d\tht d\bar\tht  D\bar\phi \bar D\phi} + q\int{dt d\tht d\bar\tht K'(\phi, \bar\phi)}, \label{S2m}
\ee
where $K'(\phi,\bar\phi)$ is the superpotential given by the redefined K\"ahler prepotential of Eq.(\ref{redef-K}), now in terms of the chiral and antichiral superfield coordinates of the particle, $\phi$ and $\bar\phi$: 
\begin{eqnarray}
          \phi(t,\tht,\bar\tht) = z(t) + \tht\xi(t) - i\tht\bar\tht\dot z(t)  \\
          \bar\phi(t,\tht,\bar\tht) = \bar z(t) - \bar\tht\bar\xi(t) + i\tht\bar\tht\dot{\bar{z}}(t) , 
\end{eqnarray}
satisfying $\bar D\bar\phi= D\phi=0$, and with $z(t)=x(t)+iy(t)$ being the complex variable representing the real coordinates $x(t)$ and $y(t)$ of the particle, and $\xi(t)$ its Grassmanian supersymmetric partner. The supersymmetry derivatives are defined as
\begin{eqnarray}
       D = \p_{\bar\tht} - i\tht\p_t  \\
       \bar D = \p_\tht - i\bar\tht\p_t .
\end{eqnarray}

The superaction (\ref{S2m}) reads in components as $S_2\equiv\int{dt L_2}$, with
\begin{eqnarray}
     L_2 & = & \frac{M{\dot{\vec{x}}}^2}{2} - i\frac{M}{8}(\dot\xi\bar\xi + \dot{\bar\xi}\xi) + q\dot{\vec x}\cdot\vec A
                     - g\dot{\vec x}\cdot\widetilde{\vec E} + \nonumber \\*
           &   & {} + \frac{q}{8}[\xi,\bar\xi]B + \frac{g}{8}[\xi,\bar\xi](\vec\na\cdot\vec E), \label{L2m}
\end{eqnarray}
which, as expected, is the same result that would be obtained if one had started with a minimal superaction, i.e., Eq.~(\ref{S2m}) with $K(\phi,\bar\phi)$ replacing $K'(\phi,\bar\phi)$, and the non-minimal prescription had been implemented only after the corresponding splitting in components, by means of Eqs.~(\ref{redef-A}) and (\ref{redefB}).
Moreover, this Lagrangian is identical to the N=1 case, Eq.~(\ref{L1}), provided the identification $\psi = \frac{\sqrt{M}}{2}\xi$ is made. Thus, all the quantization procedure carried out after Eq.~(\ref{L1}) may be repeated, yielding the same results and attesting, in a superfield description, the fact that in (2+1) dimensions the Pauli equation possesses, rather than an N=1-, an N=2-SUSY\cite{g2sqm} (note that this conclusion is valid independently whether the coupling is minimal or non-minimal).

It should be noticed that the extension from N=1- to N=2-SUSY carried out here is simply due to a well-established result\cite{SQMbook2,g2sqm}: N=2 is the true SUSY of (2+1)-dimensional Pauli equation, in contrast to the (3+1)-dimensional case, which is just N=1-supersymmetric. Therefore, this part of the present work should not be confused with the equivalence between N=1- and N=2-SQM as shown in Ref.~\cite{CoGiKi04} even for the (3+1)-dimensional Pauli equation. In the latter case, the second supercharge is non-local (since it involves explicitly the parity operator). Indeed, a sensible question that remains to be investigated in detail is the transition from such a non-local supercharge of the (3+1)-dimensional Pauli equation to a local one, when the dynamics is restricted to be (2+1)-dimensional; or, equivalently, when the (3+1)-dimensional magnetic field is considered to depend only on $x$ and $y$ and to be parallel to the $z$-axis (there is also a third possibility: when the magnetic field has a definite space parity, $\vec{B}(-\vec{r}) = \pm \vec{B}(\vec{r})$; see Ref.~\cite[p.110]{SQMbook2} or Ref.~\cite{SQMreview0}).

\section{\label{concl}Discussion and conclusions}

Here, it has been shown the possibility, expressed by Eq.~(\ref{Phi-SUSY}), for SUSY to be kept even with an electric field applied, provided a non-minimal coupling scheme holds.
Moreover, since this work deals with planar physics, it may suggest a possible application 
to the quantum Hall effect (QHE)\cite{FQHE1,FQHE2,FQHE3,FQHE4}.
Indeed, such a possibility was already pointed out in Ref.~\cite{PLA}, where a parallel was made between the particle with charge and magnetic moment as described in Section~\ref{nmin} and the composite fermion of Jain's model for the fractional QHE\cite{Jain1,Jain2,Jain3,Jain4}. Now, assuming the validity of such a parallel, the present work brings a SUSY to the system of composite fermions. Another result is Eq.~(\ref{g-crit}), a condition also obtained in field theoretical works (with other interpretations) and which here guarantees the gyromagnetic ratio to be equal to its standard value, two. All the calculations are made in superfield formulation.

Finally, a more general possibility for the interacion will be discussed, in which the following terms are added to the superaction (\ref{S2m}):
\be
       \int{dt d\tht\: \Gamma(\phi)} + \int{dt d\bar\tht\:\bar{\Gamma}(\bar\phi)}, \label{SF}
\ee
where $\Gamma(\phi)$ and its complex conjugate $\bar{\Gamma}(\bar\phi)$ are necessarily Grassmann external fields, in order to the action be bosonic. The corresponding components added to the Lagrangian (\ref{L2m}) are:
\be
       \xi \Gamma'(z) - \bar\xi \bar{\Gamma}'(\bar z),                 \label{L2mF}
\ee
where the primes in the $\Gamma$s now stand for differentiation with respect to the argument.

The (pseudo-)classical external field $\Gamma'(z)$, although not quantized, 
must anticommute with $\xi$ 
as well as with itself. Therefore, it may be represented also by a $2\times 2$-matrix. These anticommutation requirements, however, impose such severe restrictions on the matrix $\Gamma'(z)$ that, under the quantization procedure mentioned in Section~\ref{N=1}, the contribution of the terms (\ref{L2mF}) to the Hamiltonian is zero. To bypass such an obstruction, it is reasonable to open the possibility of a different matrix representation for the Grassmannian coordinates, $\xi$ and $\bar\xi$, and their corresponding momenta. Indeed, adopting, for example, the $4 \times 4$ representation below:
\be
\xi=\sigma_+\otimes  \openone_{2\times 2} \equiv \left( \begin{array}{cccc}
                                                                            0 & 1 & 0 & 0 \\
                                                                            0 & 0 & 0 & 0 \\
                                                                            0 & 0 & 0 & 1 \\
                                                                            0 & 0 & 0 & 0  \end{array} \right)  
\ee
and 
\be
       \Gamma'(z) \equiv \left( \begin{array}{cccc}  
                                              f(z) & g(z) & h(z) & i(z) \\
                                              j(z) & k(z) & l(z) & m(z) \\
                                             n(z) & o(z) & p(z) & q(z) \\
                                             r(z) & s(z) & t(z) & u(z)
\end{array} \right),
\ee
the same anticommutation requirements for $\Gamma'(z)$ lead to the following total Hamiltonian: 
\begin{eqnarray}
H_2 & = & \frac{\left(\vec p - q\vec A + g\widetilde{\vec E}\right)^2}{2M} 
                 - \frac{qB}{2M}\si_3\otimes \openone_{2\times 2} + \nonumber \\*
       &    &  {} - \frac{g(\vec\nabla\cdot\vec E)}{2M}\si_3\otimes \openone_{2\times 2} + \frac{2}{\sqrt{M}}G(z,\bar z) ,          
\end{eqnarray}
where
\be
       G(z,\bar z) = \left(  \begin{array}{cccc}
           0               &          -f(z)        &                0                 &         -h(z)   \\
-\bar{f}(\bar z)      &             0         &        \frac{\bar{f}^2(\bar z)}{\bar{h}(\bar z)}   &            0     \\        
           0               &      \frac{f^2(z)}{h(z)}   &               0                 &          f(z)    \\
 -\bar{h}(\bar z)    &             0         &       \bar{f}(\bar z)      &           0
                                          \end{array}  \right).
\ee

Notice that this interaction mixes the four components of the wave function, contrary to the original Hamiltonian. 
The Grassmann fields $\Gamma'(z)$ and $\bar{\Gamma}'(\bar z)$ may be interpreted as photino-type (pseudo-)classical external fields, in the same way as the electromagnetic prepotential $K'(z,\bar z)$ (or the potentials $\Phi$ and $\vec A$) is usually considered as a photon-type classical external field. The motivation for adopting 4-component wave functions  is similar to what happens in (2+1)D field theories with massive fermions, when one is forced to introduce 4-component spinors, rather than 2-component ones, in order that the mass term be compatible with parity symmetry. This is a peculiar feature of planar theories.

\begin{acknowledgments}
The authors thank Germano Monerat for fruitful discussions at an early stage of this work. RCP thanks Marcelo Carvalho for helpful discussions. LPGA expresses his gratitude to CNPq-Brazil for his Graduate fellowship.
\end{acknowledgments}

%\bibliography{SQM-f}

\end{document}